\documentclass[aps,prx,twocolumn,notitlepage,showpacs,superscriptaddress,10pt]{revtex4-2}%
\usepackage{graphicx}
\usepackage{amsmath}
\usepackage{amssymb}
\usepackage{color}
\usepackage{amsfonts}%
\usepackage[caption=false]{subfig}

\usepackage{color}
\usepackage[dvipsnames]{xcolor}
\usepackage[colorlinks,bookmarks=false,citecolor=darkblue,linkcolor=red,urlcolor=blue]{hyperref}
\usepackage[capitalize]{cleveref}

\definecolor{darkred}{rgb}{0.7,0.0,0.0}

\definecolor{darkblue}{rgb}{0,0.02,0.45}

\definecolor{darkgreen}{rgb}{0.02,0.45,0.0}

\definecolor{violet}{rgb}{0.8,0.2,0.6}

\providecommand{\U}[1]{\protect\rule{.1in}{.1in}}

\newcommand{\NRC}{MA$_2$NaRuCl$_6$}
\newcommand{\RB}{MA$_3$Ru$_2$Br$_9$} 
\newcommand{\AW}{A$_2$WCl$_6$} 
\newcommand{\MAW}{MA$_2$WCl$_6$} 
\newcommand{\CsR}{Cs$_2$RuCl$_6$} 
\newcommand{\CsW}{Cs$_2$WCl$_6$} 

\begin{document}
\title{Microscopic origin of temperature-dependent magnetism in spin-orbit-coupled transition metal compounds}


\author{Ying Li}\thanks{yingli1227@xjtu.edu.cn}
\affiliation{MOE Key Laboratory for Nonequilibrium Synthesis and Modulation of Condensed Matter, School of Physics, Xi'an Jiaotong University, Xi'an 710049, China}
\author{Ram Seshadri}\
\affiliation{Materials Department, University of California, Santa Barbara, CA 93106, USA}
\author{Stephen D. Wilson}\
\affiliation{Materials Department, University of California, Santa Barbara, CA 93106, USA}
\author{Anthony K. Cheetham}\
\affiliation{Materials Research Laboratory, University of California, Santa Barbara, CA 93106, USA}
\affiliation{Department of Materials Science and Engineering, National University of Singapore, Singapore 117575}
\author{Roser Valent{\'\i}}\thanks{valenti@itp.uni-frankfurt.de}
\affiliation{Institut f\"ur Theoretische Physik, Goethe-Universit\"at Frankfurt,
Max-von-Laue-Strasse 1, 60438 Frankfurt am Main, Germany}
\date{\today}

\begin{abstract}
A few $4d$ and $5d$ transition metal compounds with various electron fillings were recently found to exhibit magnetic susceptibilities $\chi$ and magnetic moments that deviate from the well-established Kotani model. This model has been considered for decades to be the canonical expression to describe the temperature dependence of magnetism in systems with non-negligible spin-orbit coupling effects. 
In this work, we uncover the origin of such discrepancies and  determine the applicability and limitations of the Kotani model by calculating the temperature dependence of the magnetic moments of a series of $4d$ (Ru-based) and $5d$ (W-based) systems at different electron fillings.
For this purpose, we perform exact diagonalization of \textit{ab initio}-derived relativistic multiorbital Hubbard models on finite clusters and compute  their magnetic susceptibilities. Comparison with experimentally measured magnetic properties indicates that contributions such as a temperature independent $\chi_0$ background, crystal field effects, Coulomb and Hund's couplings, and intersite interactions -- not included in the Kotani model -- are specially crucial to correctly describe the temperature dependence of $\chi$  and magnetic moments at various electron fillings in these systems. Based on our results, we propose a generalized approach beyond the Kotani model to accurately describe their magnetism.

\end{abstract}
\maketitle
\par

\begin{figure}
\center
\includegraphics[angle=0,width=\linewidth]{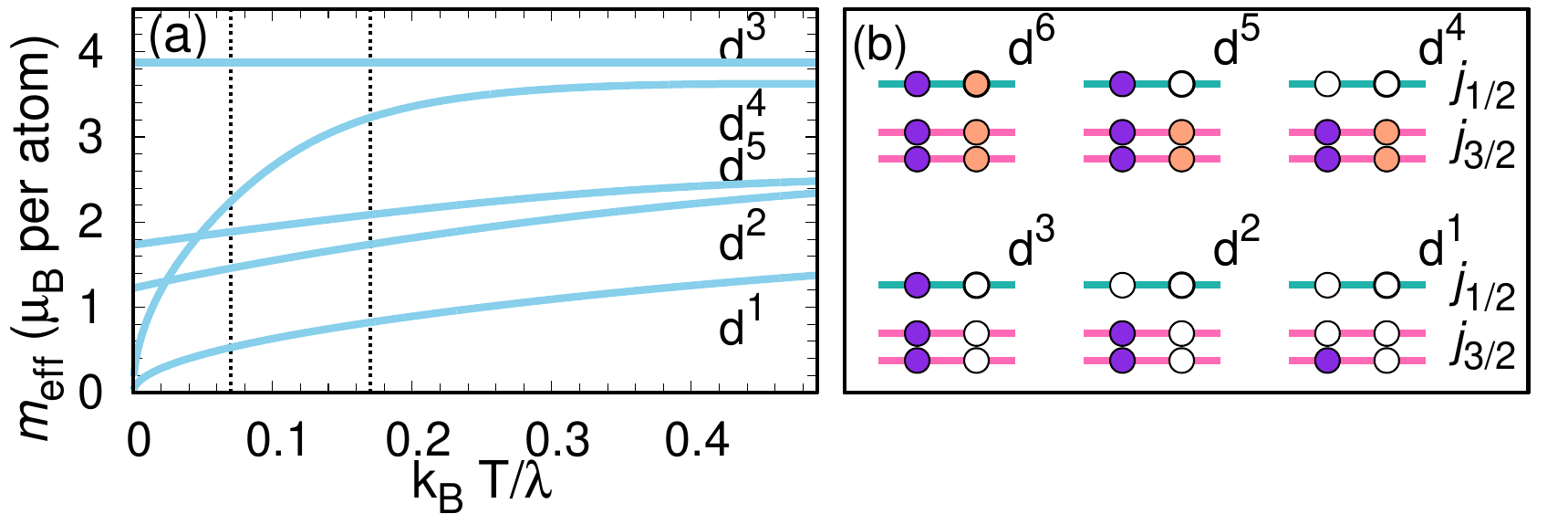}
\caption{(a) Kotani plot of $m_{\rm eff}$ for various fillings of $t_{2g}$ orbitals. The dotted vertical lines at $k_{\rm B}$T/$\lambda \sim$ 0.07 and 0.17 eV indicate T = 300 K when $\lambda$ = 0.15 eV (4$d$) and 0.37 eV (5$d$), respectively. (b) Schematic ground state diagrams for each filling in the relativistic $j_{1/2}$ and $j_{3/2}$ basis.}
\label{fig:kotani}
\end{figure}

{\it Introduction.-} 
The magnetism in spin-orbit coupled transition-metal-based compounds has been at the focus of intensive studies in recent years, not only due to their relevance for such phenomena as
frustrated magnetism~\cite{Normand2009,Balents2010, Zhou2017, Savary2017,Broholm2020}, 
 Kitaev spin liquid phases~\cite{Kitaev2006,Jackeli2009,Chaloupka2013, Witczak-Krempa2014,Rau16,Schaffer2016,WinterReview,Cao2018,kaib2019kitaev,Takagi2019,Trebst2022}, fractionalized excitations~\cite{Savary2017, Zhou2017, Hermanns2018,Chern2019,Knolle2015}, or mutipolar physics~\cite{Chen2010, Erickson2007, Lu2017, Hirai2020, Frontini2024}, but also for their role in, for instance, $4d$ and $5d$  hybrid halides, which are being discussed as relevant materials for optoelectronic purposes~\cite{Kojima2009, Green2014, Burschka2013, Im2014, Vishnoi2020}.
Due to the presence of spin-orbit coupling, the usual Curie-Weiss formula to explain the temperature-dependent susceptibility in these systems does not apply. To overcome this problem 
Kotani~\cite{Kotani1949} proposed a temperature-dependent expression for
magnetic moments [see Fig.~\ref{fig:kotani} (a)], which successfully explains the behavior of the magnetic susceptibility in a few $4d$ and $5d$ systems.  Some examples are ruthenium halides with perovskite-related structures such as A$_2$RuX$_6$ (A = K, Cs, Rb; X = Cl, Br)~\cite{Johannesen1963, Earnshaw1961, Vishnoi2021}, and a few cases of Ru-based hybrid halides~\cite{Lu2018,Kotani1990}. 
In some other examples, including Ru-based hybrid halides \NRC(MA = CH$_3$NH$_3$), \RB, and vacancy-ordered double perovskites \AW~(A = Cs, Rb, MA)~\cite{Kennedy1963, Liu2022, Morgan2023}, the temperature-dependent magnetic susceptibility deviates strongly from the Kotani prediction, challenging this well-established model for spin-orbit-coupled transition-metal complexes.

We recently considered extensions of Kotani's expression to describe the magnetism in a few cases of spin-orbit-coupled magnets, such as the iridium and ruthenium-based $d^5$ Kitaev systems Na$_2$IrO$_3$, Li$_2$IrO$_3$, RuCl$_3$, and RuBr$_3$~\cite{Li2021, Kaib2022}, as well as  
iridium-based mixed-valence hexagonal perovskites~\cite{Li2020} by including trigonal distortions~\cite{Kamimura1956} on a single-site description of the octahedral environment of the transition-metal $d$ states and derived a modified Curie-Weiss formula in terms of a temperature-dependent magnetic moment. Whereas the temperature dependence of the average magnetic moment in the above $d^5$ Kitaev systems is not affected by small trigonal distortions and follows the Kotani model, the components
of the magnetic moments do not follow the Kotani formula~\cite{Li2021, Kaib2022}. For the $4d$- and $5d$-based  mixed-valence systems with dimerized structures, the Kotani expression
fails completely to reproduce their temperature-dependent magnetism~\cite{Li2020}. The question is therefore, why does the Kotani formula describe quite accurately the temperature dependence of the magnetic susceptibility and magnetic moments in some spin-orbit coupled systems and it does not for others?

In the present work we uncover the origin of this discrepancy. Specifically we resolve i) the role of electronic fillings shown in Fig.~\ref{fig:kotani} (b), ii) the 
sensitivity of the magnetic moment components to the crystal-field distorsions, iii) the role of Hund's coupling $J_H$, specially  in $d^2$ systems, iv) the role of intersite hoppings and interactions in dimer systems, and v) the role of $t_{2g}-e_g$ crystal fields~\cite{Pradhan2024}.  
For this purpose,  we investigate the magnetic behavior of a few representative systems including \NRC~($d^5$), \RB~($d^5$), \CsR ~($d^4$), and \AW~($d^2$) by performing  exact diagonalization of \textit{ab initio}-derived relativistic multiorbital Hubbard models on finite clusters, computing their
magnetic susceptibilities and extracting general rules for the validity range of the Kotani model.

\begin{figure}
\includegraphics[angle=0,width=\linewidth]{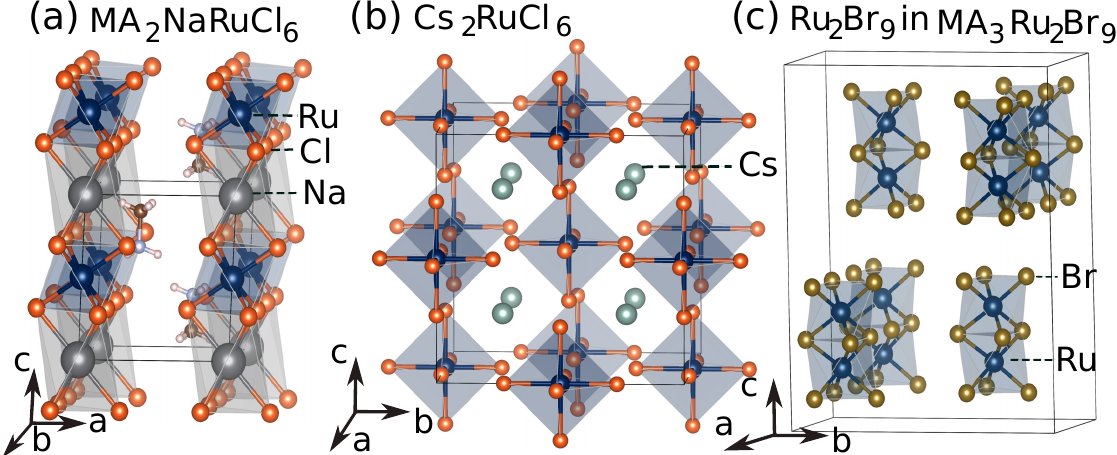}
\caption{Crystal structure of (a) \NRC~(b) \CsR~and (c) Ru$_2$Br$_9$ face-shared bioctahedral dimer in \RB. 
}
\label{fig:structure}
\end{figure}

{\it Methods.-} To analyze the systems' magnetism we consider the improved Curie-Weiss expression for the diagonal components of the susceptibility~\cite{Li2021}, 
\begin{align}
\chi^{\alpha}(T) \approx &  \ \chi_0^\alpha  + \ 
\frac{N_A[m_{\rm eff}^\alpha(T)]^2}{3k_B(T-\Theta^\alpha)},
\label{eq:sus}
\end{align}
where $\alpha$, $T$, and $\chi^{\alpha}_0$ indicate the field direction, temperature, and temperature-independent background contributions, respectively. $N_A$, $k_B$ and $\Theta^{\alpha}$ are Avogadro, Boltzmann and Weiss constants, respectively.  $m_{\rm eff}$ denotes the effective magnetic moment. 
In systems with weak inter-site interactions, the experimentally measured susceptibility is usually mapped to a magnetic moment 
 \begin{align}
\tilde{m}^{\alpha}_{\rm eff}(T) \approx \sqrt{\frac{3k_BT}{N_A}\chi^{\alpha} (T)}.
\label{eq:mu}
\end{align}
Here, we denote $\tilde{m}^{\alpha}_{\rm eff}(T)$ = $m^{\alpha}_{\rm eff}$(T) only when inter-site hoppings, $\chi_0$, and $\Theta^{\alpha}$ are set to zero. In our microscopic approach, we use Eq.~\ref{eq:mu}~\cite{Li2020, Li2021} to obtain the magnetic moments
where the susceptibility is calculated from:
\begin{equation}
\chi^{\alpha}(T) = \chi_0 + N_A k_BT\left(\frac{1}{Z} \frac{\partial^2 Z}{\partial H_{\alpha}\partial H_{\alpha}}\right).
\label{eq:chi}
\end{equation}
  $H$ is the external magnetic field and $Z$ is the partition function 
$Z = \sum_n e^{-E_n/k_BT}$
with $E_n$ being the eigenenergies obtained from exact diagonalization on finite clusters of the Hamiltonian $\mathcal{H}_{\rm tot}$ including the kinetic hopping term ($\mathcal{H}_{\rm hop}$), crystal-field splitting ($\mathcal{H}_{\rm CF}$), spin-orbit coupling ($\mathcal{H}_{\rm SO}$), and Coulomb interaction ($\mathcal{H}_U$) (see Suppl. Material). The corresponding parameters are hopping parameters $\textbf{t}$, crystal-field parameters ($\Delta$), spin-orbit coupling parameter $\lambda$, Coulomb repulsion  $U$, and Hund's coupling $J_{\rm H}$.
$\textbf{t}$ and $\Delta$ were extracted using the Wannier function projection formalism~\cite{aichhorn2009,Foyevtsova2013,ferber2014} from the electronic structure of density functional theory with the full-potential-linearized-augmented-plane-wave basis(LAPW)~\cite{Wien2k} (see Suppl. Material).

\begin{figure}
\includegraphics[angle=0,width=1\linewidth]{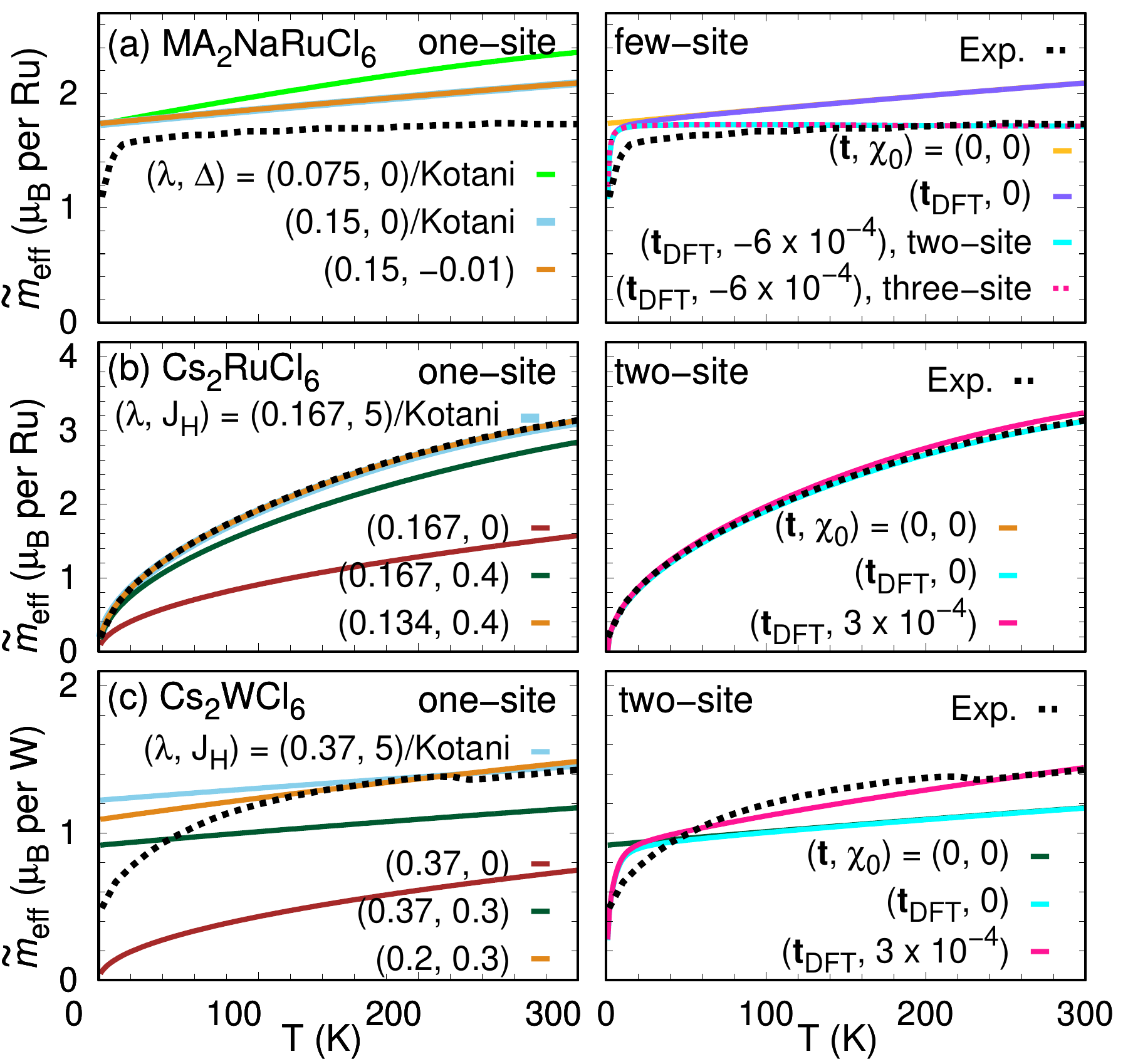}
\caption{Temperature dependence of the average magnetic moments for (a) \NRC~($d^5$) (b) \CsR~($d^4$) and (c) \CsW~($d^2$) from one-site, two-site, and three-sites cluster calculations with various spin-orbit coupling $\lambda$, trigonal crystal-fields $\Delta$, Hund's couplings $J_{\rm H}$, and hopping parameters \textbf{t} in eV as well as $\chi_0$ in emu per mol.}
\label{fig:momd}
\end{figure}

{\it Results.-} 
The dominant crystal information and hopping parameters (summarized in Table~\ref{tab:hop}) are discussed here and the details are in the Supplemental Material. Due to the octahedral environment, the transition-metal $d$ states split into  $e_g$ and $t_{2g}$ states. We first consider the $t_{2g}$-only model widely used in these systems.

\begin{table}[]
    \centering\def\arraystretch{1.1}
       \caption{Largest hopping parameters in meV extracted from DFT calculations for $d^5$ (\NRC, \RB), $d^4$ (Cs$_2$RuCl$_6$) and $d^2$ (Cs$_2$WCl$_6$) systems.}
    \begin{ruledtabular}
    \begin{tabular}{cccc}
     \NRC & \RB & Cs$_2$RuCl$_6$ & Cs$_2$WCl$_6$\\     
     -39.1 & -338.6 & -38.4 & -46.6 \\
    \end{tabular}
    \end{ruledtabular}
    \label{tab:hop}
\end{table}
We start with a $d^5$ system \NRC~ which crystallizes in the space group $P\overline{3}m$ and contains chains of face-sharing RuCl$_6$ and NaCl$_6$ octahedra along $c$ [Fig.~\ref{fig:structure} (a)]. In order to explain the experimentally observed effective magnetic moment
for \NRC~\cite{Vishnoi2020},  which deviates from the ideal Kotani model~\cite{Kotani1949}, we apply the method described above to single-site, two-site and three-site Ru clusters. We used $U$ = 2.49 eV and $J_{\rm H}$ = 0.4 eV following the cRPA values of $\alpha$-RuCl$_3$~\cite{Kaib2022}. The results for $\tilde{m}_{\rm eff}$ are shown in Fig.~\ref{fig:momd} (a).  $\tilde{m}_{\rm eff}$ = $m_{\rm eff}$ for the one-site case while $\tilde{m}_{\rm eff}$ $\neq$ $m_{\rm eff}$ for the few-sites case with nonzero \textbf{t} and $\chi_0$. We observe that
the experimental results~\cite{Vishnoi2020} show lower values of the effective magnetic moment in comparison with the ideal Kotani behavior ($\lambda$ = 0.075 eV), and the slopes at low and high temperatures do not fit either. Increasing the value of $\lambda$ to 0.15 eV, which is comparable to values for other Ru-based systems like $\alpha$-RuCl$_3$, reduces the slope of the calculated magnetic moment towards the experimental slope. Including the effect of the trigonal crystal-field distortion $\Delta$ = -0.01 eV does not affect the average magnetic moment, but strongly affects the direction-dependent magnetic moments (see Supplemental Material).


Including the contribution of all hopping parameters in the two-site calculation (\textbf{t} = \textbf{t}$_{\rm DFT}$) introduces a sharp reduction of the susceptibility at low temperatures. Actually, the experimentally measured susceptibility also includes a temperature-independent weak contribution $\chi_0$, originating from core diamagnetism and Van-Vleck paramagnetism~\cite{Nag2017,Orbach1966}. With $\chi_0$ = -6 $\times$ $10^{-4}$ emu/mol the effective magnetic moment becomes flat for a range of temperatures between 30 K and 300 K, close to the experimental magnetic moment.  
We conclude that
the deviation of the experimental magnetic moment from the Kotani plot in $d^5$ system can be attributed to both (i) a temperature-independent $\chi_0$ and (ii) inter-site interactions. Results for two-site and three-site clusters are very close so that we  only consider two-site clusters in the following cases.

We now proceed with the $d^4$ case. The $4d$-based hybrid vacancy-ordered double perovskite \CsR~crystallizes in the $Fm\overline{3}m$ space group and consists of isolated RuCl$_6$ octahedra bound electrostatically by Cs cations [see Fig.~\ref{fig:structure} (b)]. 
The largest hopping parameters is -38.4 meV. The results are displayed in Fig.~\ref{fig:momd} (b). The Kotani plot can be reproduced by considering large unphysical $J_{\rm H}$ (around 5 eV) with $\lambda$ = 0.167 eV, or, alternatively choosing $\lambda$ = 0.134 eV, $J_{\rm H}$ = 0.4 eV. In fact, various combination of $J_{\rm H}$ and $\lambda$ could reproduce the experimental observations. We also performed two-site cluster calculations with $J_{\rm H}$ = 0.4 eV and find that including $\chi_0$ and hopping parameters modify only slightly the results. 

\begin{figure}
\center
\includegraphics[angle=0,width=\linewidth]{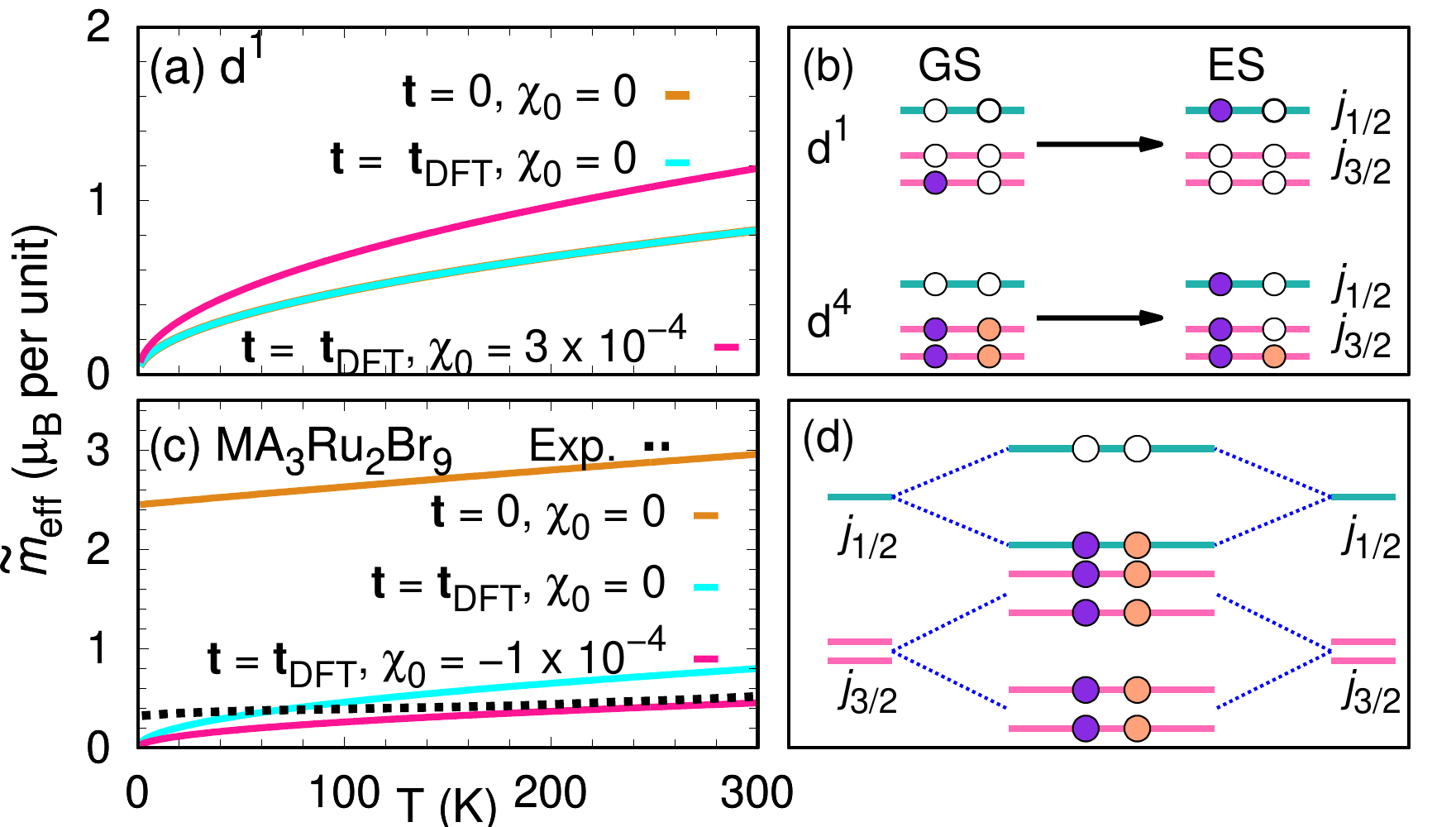}
\caption{The average magnetic moments for (a) $d^1$ system and (c) dimer system \RB~with two-site cluster calculations. 
Schematic diagrams of (b) ground states and first excited states for $d^1$ and $d^4$ systems, as well as (d) the ground state of the dimer system~\RB~.}
\label{fig:d1dimer}
\end{figure}

In contrast to $d^4$ systems, $d^2$ systems, such as \CsW, with similar crystal structures to \CsR~ by replacing Ru with W, display a different behavior. We used $U$ = 1.7 eV following values previously used for Ir-based systems~\cite{Winter2016}. The results are shown in Fig.~\ref{fig:momd} (c). The Kotani plot with $\lambda$ $\sim$ 0.37 eV~\cite{Morgan2023} can only reproduce the slope of the experimental data by considering large unphysical $J_{\rm H}$ (around 5 eV) where the triplet state is then the ground state in Kotani's perturbation theory. For $J_{\rm H}$ = 0, the ground state is the singlet state with zero magnetic moment at 0 K. In $5d$ systems usual values of  $J_{\rm H}$  are around 0.3 eV~\cite{Winter2016}, and the ground state is a mixing of triplet and singlet states. The magnetic moment is therefore between the above two limits. When decreasing $\lambda$ to 0.2 eV, the contribution of the triplet states becomes larger, leading to larger magnitudes of the magnetic moments. The two tuning parameters for the magnetic moments are therefore Hund's and spin-orbit coupling strengths. However, different from the $d^4$ case, the Kotani plot cannot reproduce the experimental data with a combination of $\lambda$ when assuming physically reasonable values of $J_{\rm H}$ = 0.3 eV. For two-site clusters, the magnitude of the effective magnetic moment is sharply reduced at low temperatures when $\mathbf{t} = \mathbf{t}_{\rm DFT}$. Including a positive $\chi_0$ brings a downturn in the magnetic moments, close to the experiment. Our method also reproduces the experimental observations for other $d^2$ systems Rb$_2$WCl$_6$ and \MAW~ (see the Suppl. Material).
We conclude that in $d^2$ systems, besides the contributions of $\chi_0$ and intersite interactions, as in the $d^5$ systems, the Hund's coupling plays an important role in reproducing the experimental magnetic moments. Finally, for $d^1$ systems, inclusion of $\mathbf{t}_{\rm DFT}$, does not affect  $\tilde{m}_{\rm eff}$ while adding a nonzero $\chi_0$ significantly changes the slope [see Fig.~\ref{fig:d1dimer} (a)].

Analyzing the previous results we observe several facts.  Already at the level of the Kotani model, the dependence of $\tilde{m}_{\rm eff}$ on the electron filling is rather remarkable, as shown in Fig.~\ref{fig:kotani} (a). For $t_{2g}$ states with spin-orbit coupling, the ground state is more conveniently described in terms of the doublet $j_{1/2}$ and quadruplet $j_{3/2}$ [Fig.~\ref{fig:kotani} (b)]. Different from the atomic relativistic basis $j$ for $p$ orbitals with $L_p$ = 1, here $j_{1/2}$ and $j_{3/2}$ correspond to $L_{\rm eff}$ = -$L_p$.
 With an electron filling of  $d^5$, $m_{\rm eff}$ = 1.73 $\mu_B$ at 0 K which corresponds to one hole in the relativistic $j_{1/2}$ basis, while with an electron filling of $d^2$, $m_{\rm eff}$ = 1.22 $\mu_B$ at 0 K corresponds to a triplet state (in reality there is a reduction of this value due to mixing with a singlet state). For these two cases, inclusion of small intersite hoppings  significantly reduces $\tilde{m}_{\rm eff}$ at low temperatures, as demonstrated above.
 
In contrast, at electron fillings of $d^1$ and $d^4$, intersite hoppings seem to not affect the effective magnetic moments.
At these fillings, the Kotani model predicts $m_{\rm eff}$ = 0 at 0 K. 
These two cases are affected differently by $\chi_0$. For $d^4$ systems, the effective magnetic moment changes sharply at values below $k_{B}T/\lambda \sim$ 0.2 and is less influenced by other interactions while the curves for $d^1$ (also all other fillings) have a smaller slope and are more sensitive to the consideration of further interactions. This could be explained by analyzing the role of the excited states [see Fig.~\ref{fig:d1dimer} (b)]. In $d^1$, there are only two energy levels, the quadruple degenerate ground state $j_{3/2}$ and the doubly degenerate excited state $j_{1/2}$ (magnetic moment 1.73 $\mu_{\rm B}$), the energy difference is 3$\lambda$/2 $\sim$ 0.22 eV ($\lambda = 0.15$ eV). At high temperatures, the spin and orbital degrees of freedom are not coupled, therefore the total magnetic moment is $\sqrt{l(l+1)+4s(s+1)} = \sqrt{5}$ ($s=\frac{1}{2}, l=1$). The magnetic moment changes slowly from 0 to values around $\sqrt{5}$. In $d^4$, the energy difference of the ground state and first excited state is $-\frac{1}{2} (5J_H - \sqrt{25J_H^2+10J_H\lambda+9\lambda^2}) \sim$ 0.096 eV ($\lambda \sim$ 0.15 eV, $J_{\rm H} \sim$ 0.4 eV). There are more available energy levels and the highest magnetic moment at high temperatures  is $\sqrt{10}$ ($s = 1, l=1$). Therefore, the magnetic moment increases dramatically from 0 to $\sqrt{10}$ in a small temperature region. 
For $d^3$ systems, the effective magnetic moment is constant in the Kotani's plot [see Fig.~\ref{fig:kotani} (a)] because the orbital moment is quenched. However, with increasing spin-orbit coupling, in the $j_{\rm eff}$ basis, the total magnetic moment reduces at lower temperatures. This reduction has been observed experimentally in NaOsO$_3$~\cite{Shi2009} and SrTcO$_3$~\cite{Rodriguez2011}, and theoretical investigated in Ref.~\cite{Matsuura2013}.

All systems considered so far have weak inter-site hoppings below 65 meV.  
In what follows we consider  $d^5$ \RB~which is formed by Ru$_2$Br$_9$  dimers comprising two face-sharing RuBr$_6$ octahedra [see Fig.~\ref{fig:structure} (c)]. The largest nearest neighbor intra-dimer hopping is -338.6 meV. In order to explain the experimental observed effective magnetic moment, which shows a dramatically low, temperature-independent moment compared to the predicted Kotani behavior, we calculate the temperature-dependent magnetic moments using two-site clusters shown in Fig.~\ref{fig:d1dimer} (c). Including the intra-dimer hopping parameters, the strong binding splits the single site relativistic $j_{1/2}$ and $j_{3/2}$ basis to form molecular bonding and antibonding energy levels [see Fig.~\ref{fig:d1dimer} (d)]. The ten electrons and two holes form a singlet $j_{\rm dim}$ = 0 state, similar to the case of Ba$_3$CeIr$_2$O$_9$~\cite{Revelli2019}, leading to a significant reduction of the magnetic moments as shown in Fig.~\ref{fig:d1dimer} (c). Different from experiment, the magnetic moments approach zero at 0 K.
These deviations  have been suggested to originate from the intra- versus inter-dimer interactions in $d^4$ systems Ba$_3$ZnIr$_2$O$_9$~\cite{Nag2016}, Ba$_3$CdIr$_2$O$_9$~\cite{Khan2019}, and Ba$_3$MgIr$_2$O$_9$~\cite{Nag2018}. In our case, inter-dimer interactions and possible impurity effects are not included in our calculations.

\begin{figure}
\center
\includegraphics[angle=0,width=\linewidth]{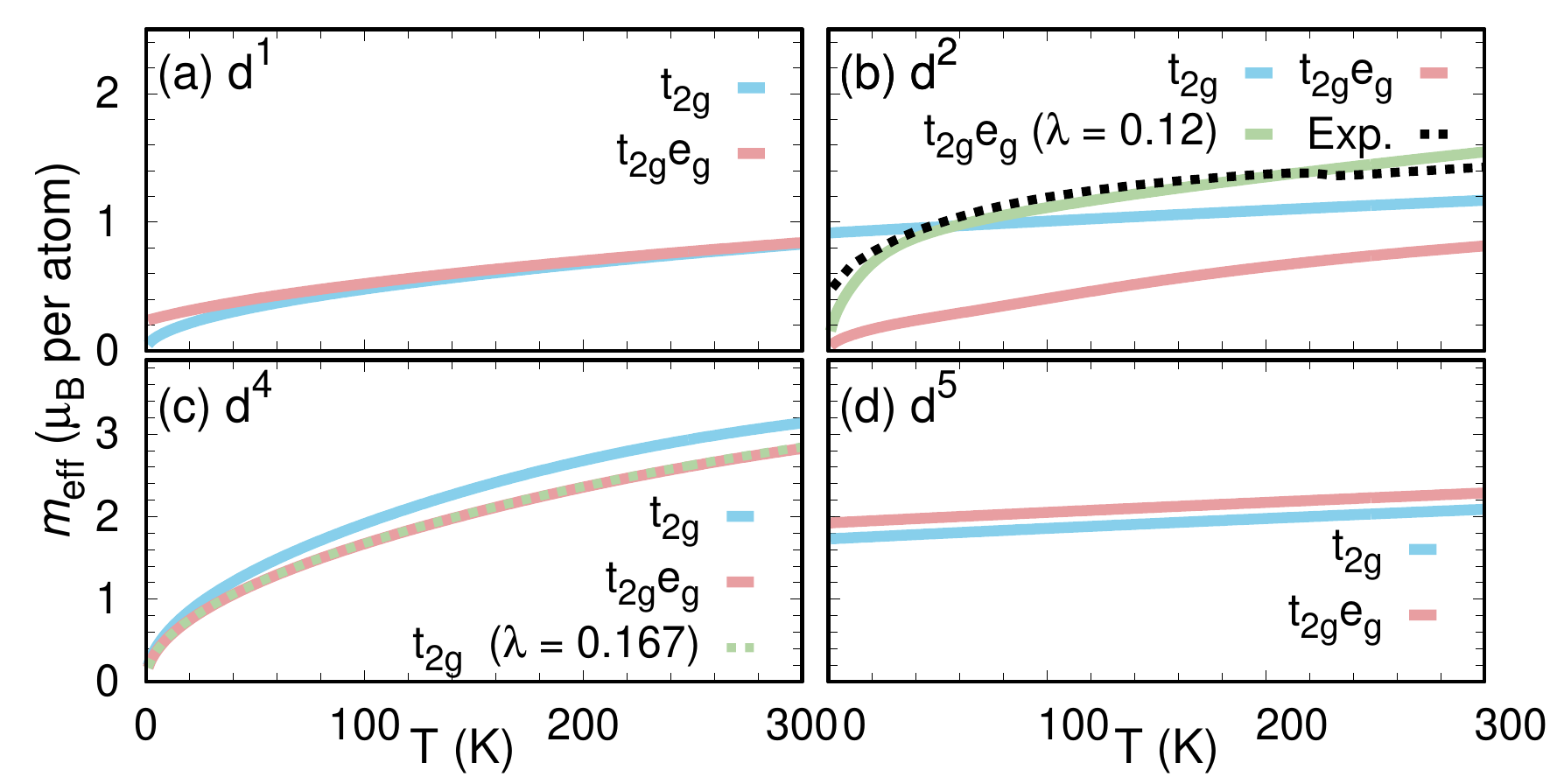}
\caption{Temperature dependence of the average magnetic moment for (a) $d^1$ (b) $d^2$ (c) $d^4$ and (d) $d^5$ including $t_{2g}$ and $e_g$ orbitals.}
\label{fig:tegeg}
\end{figure}

From the above results, we observe that in a $t_{2g}$-only model, the magnetic moments for single site are determined by i) the ground state at 0 K and ii) the energy of the excited states. We now discuss whether including the $e_g$ orbitals suggested in Ref.~\cite{Stamokostas2018, Pradhan2024} affect these two factors. The results are displayed in Fig.~\ref{fig:tegeg} for various fillings ($d^1$, $d^2$, $d^4$, and $d^5$) with $t_{2g}$-$e_g$ crystal fields around 3 eV. 
For $d^1$ systems [Fig.~\ref{fig:tegeg} (a)], we find that consideration of $e_g$ states induces a non-zero magnetic moment at 0 K  of the order of 0.2 $\mu_B$, similar to the experimental value for Cs$_2$TaCl$_6$ and Rb$_2$TaCl$_6$~\cite{Ishikawa2019}.
In our calculations the small magnetic moments originate from  non-zero  $j_{1/2}$ states arising from matrix elements between $e_g$ and $t_{2g}$ states. For $d^2$ systems [Fig.~\ref{fig:tegeg} (b)] as in Cs$_2$WCl$_6$, inclusion of $e_g$ states splits the five degenerate $d$ states to a doublet ground states and triplet excited states, strongly reducing the magnetic moment at 0 K
[red curve in Fig.~\ref{fig:tegeg} (b)]. The energy difference between doublet and triplet is 0.03 eV. By decreasing $\lambda$ to 0.12 eV, the energy of the first excited states is reduced and the magnetic moment increases dramatically [green curve in Fig.~\ref{fig:tegeg} (b)], similar to the experimental observations. Different from Ref.~\cite{Pradhan2024}, where only the Kanamori model and the first excited state energy gap were used to calculate the susceptibility, our model includes full Coulomb interaction matrices with all Slater parameters as well as all excited states. For $d^4$  and $d^5$ fillings, inclusion of $e_g$ states changes minimally the temperature dependence of the magnetic moments.

{\it Conclusions.-} 
Summarizing, 
we investigated the magnetic behavior of a few representative systems including \NRC~($d^5$), \RB~($d^5$), \CsR ~($d^4$), and \AW~($d^2$) by performing  exact diagonalization of \textit{ab initio}-derived relativistic multiorbital Hubbard models on finite clusters. We computed their magnetic susceptibilities and extracted general rules for the validity range of the Kotani model. We find that the slope of the effective magnetic moment is crucially dependent on the ground and excited state configurations at the various electronic fillings. In the single-site description, crystal field distortions, spin-orbit coupling, Hund's coupling, $t_{2g}$-$e_g$ orbitals, and $\chi_0$, all contribute to the temperature dependence of the effective magnetic moment in a delicate balance, depending on the electronic filling. While the Kotani model contains contribution of spin-orbit coupling and Hund's coupling,
often non-physical values need to be considered to find an agreement with experiment. 
Furthermore, for systems with strong intra-dimer interactions, such as \RB, $m_{\rm eff}$ is strongly reduced due to metal-metal bonding.

Our findings resolve the discrepancies between experimental observations and theoretical descriptions of magnetic susceptibilities and effective magnetic moments in spin-orbit-coupled transition metal complexes, 
and provide a more accurate way to understand magnetism in these systems.

{\it Acknowledgements}
We thank T. Saha-Dasgupta and A. Paramekanti for discussions.
Y.L. acknowledges support by the National Natural Science Foundation of China (Grant No.\ 12004296), the Fundamental Research Funds for the Central Universities (Grant No. xzy012023051), and HPC Platform, Xi'an Jiaotong University. R.V. acknowledges support by the Deutsche Forschungsgemeinschaft (DFG, German Research Foundation) for funding through Project No. TRR 288 --- 422213477 (project A05, B05). R.V.' s  research was supported in part by grant NSF PHY-2309135 to the Kavli Institute for Theoretical Physics (KITP). R.S. and S.D.W. thank the NSF Q-AMASE-i Quantum Foundry (DMR-1906325) for support. S.D.W. acknowledges support by DOE, Office of Science, Basic Energy Sciences under Award No. DE-SC0017752. A.K.C. thanks the Ras al Khaimah Centre for Advanced Materials for financial support.

\bibliography{ref}


\clearpage
\widetext
\appendix
\begin{center}
\textbf{\large \textit{Supplemental Material}:\\ \smallskip Microscopic origin of temperature-dependent magnetism in spin-orbit-coupled transition metal compounds
} \bigskip \bigskip
\end{center}
\twocolumngrid

\setcounter{equation}{0}
\setcounter{figure}{0}
\setcounter{table}{0}
\setcounter{page}{1}
\makeatletter
\renewcommand{\theequation}{S\arabic{equation}}
\renewcommand{\thefigure}{S\arabic{figure}}
\renewcommand{\thetable}{S\Roman{table}}

\subsection{Electronic Hamiltonian}
The Coulomb terms in the $t_{2g}$ basis of $H_{\rm tot}$ are given by:
\begin{align}
\mathcal{H}_{U}& \ = U \sum_{i,a} n_{a,\uparrow}n_{i,a,\downarrow} + (U^\prime - J_{\rm H})\sum_{i,a< b, \sigma}n_{i,a,\sigma}n_{i,b,\sigma} \nonumber \\
 &+ U^\prime\sum_{i,a\neq b}n_{i,a,\uparrow}n_{i,b,\downarrow} - J_{\rm H} \sum_{i,a\neq b} c_{i,a\uparrow}^\dagger c_{i,a\downarrow} c_{i,b\downarrow}^\dagger c_{i,b\uparrow}\nonumber \\ & + J_{\rm H} \sum_{i,a\neq b}c_{i,a\uparrow}^\dagger c_{i,a\downarrow}^\dagger c_{i,b\downarrow}c_{i,b\uparrow},
\end{align}
where $c_{i,a}^\dagger$ creates an electron in orbital $a\in\{yz,xz,xy\}$ at site $i$; $J_{\rm H}$ gives the strength of Hund's coupling, $U$ is the {\it intra}orbital Coulomb repulsion, and $U^\prime=U-2J_{\rm H}$ is the {\it inter}orbital repulsion. The one particle terms are most conveniently written in terms of:
\begin{align}
\vec{\mathbf{c}}_i^\dagger = \left(c^\dagger_{i,yz,\uparrow} \  c^\dagger_{i,yz,\downarrow} \ c^\dagger_{i,xz,\uparrow} \  c^\dagger_{i,xz,\downarrow} \ c^\dagger_{i,xy,\uparrow} \  c^\dagger_{i,xy,\downarrow}\right).
\end{align}
Spin-orbit coupling is described by:
\begin{align}
\mathcal{H}_{\rm SO}=\frac{\lambda}{2} \sum_i \vec{\mathbf{c}}_{i}^\dagger\left(\begin{array}{ccc} 0 & i \sigma_z & -i \sigma_y \\ -i \sigma_z & 0 & i\sigma_x \\ i \sigma_y & -i\sigma_x & 0\end{array} \right)\vec{\mathbf{c}}_i,
\end{align}
where $\sigma_\alpha$, $\alpha=\{x,y,z\}$ are Pauli matrices. Theis matrix is used for all materials except \RB, where two Ru atoms have two local coordiantes. For convenience, we choose as global spin quantization axis to be the crystallographic $c$-axis, but
employ local coordinates for the orbital definitions. The spin operator are then modified by the rotation from the global coordinates to the local coordinates for the orbital, similar as MO$_9$ dimers\cite{Li2020}. 
The crystal-field Hamiltonian is given by:
\begin{align}
\mathcal{H}_{\rm CF}=  \sum_i \vec{\mathbf{c}}_{i}^\dagger\left\{\mathbf{\Delta}_i\otimes \mathbb{I}_{2\times 2}\right\}\vec{\mathbf{c}}_i,
\end{align}
where $\mathbb{I}_{2\times 2}$ is the $2 \times 2$ identity matrix; $\mathbf{\Delta}_i$ is the crystal field tensor. 
The hopping Hamiltonian is most generally written:
\begin{align}
\mathcal{H}_{\rm hop}= \sum_{ij} \vec{\mathbf{c}}_{i}^\dagger \ \left\{\mathbf{T}_{ij} \otimes \mathbb{I}_{2\times 2}\right\}\ \vec{\mathbf{c}}_j,
\end{align}
with the hopping matrices $\mathbf{T}_{ij}$ defined for each bond connecting sites $i,j$. 

Including the $e_g$ for one-site cluster, the Coulomb terms are generally written as:

\begin{align}
\mathcal{H}_{U}& \ = \sum_{\alpha,\beta,\delta,\gamma} U_{\alpha\beta\gamma\delta} c_{i,\alpha,\sigma}^\dagger c_{i,\beta, \sigma^{\prime}}^\dagger c_{i,\gamma,\sigma^{\prime}} c_{i,\delta,\sigma} 
\end{align}
where $\alpha$, $\beta$, $\gamma$, $\delta$ are different $d$ orbital indices~\cite{Winter2022}. In the spherically symmetric approximation~\cite{Sugano2012, Pavarini2014}, the coefficient $U_{\alpha\beta\gamma\delta}$ are all related to the three Slater parameters $F_0$, $F_2$, $F_4$, which are expressed in terms of Kanamori parameter $U$ and $J_{\rm H}$ in $t_{2g}$ basis:
\begin{align}
F_0 & = U - \frac{13}{49} \frac{441}{79} J_{\rm H} \nonumber\\
F_2 & =  \frac{882}{79} J_{\rm H}, \nonumber \\
F_4 & = \frac{5}{4} \frac{441}{79} J_{\rm H}.
\end{align}

The formula is reduced to Kanamori formula when the number of unique orbital indices are one or two. Beyond the Kanamori formula, the Coulomb coefficients with three and four indices are significant for the small energy gap between ground state and excited states when both $e_g$ and $t_{2g}$ orbitals are included. The spin-orbit coupling is changed to:

\begin{align}
&\mathcal{H}_{\rm SO}= \nonumber \\
&\frac{\lambda}{2} \sum_i \vec{\mathbf{c}}_{i}^\dagger\left(\begin{array}{ccccc} 
0 & 0 & i\sqrt{3} \sigma_x & -i\sqrt{3} \sigma_y &  0\\
0 & 0 & i \sigma_x & i \sigma_y & -2i \sigma_z \\
-i\sqrt{3} \sigma_x &  -i \sigma_x  &    0 & i \sigma_z & -i \sigma_y \\ 
i\sqrt{3} \sigma_y&  -i \sigma_y &         -i \sigma_z & 0 & i\sigma_x \\ 
0 &   2i \sigma_z   &      i \sigma_y & -i\sigma_x & 0
\end{array} \right)\vec{\mathbf{c}}_i,
\end{align}
where the basis $\vec{\mathbf{c}}_i^\dagger$ is ($c^\dagger_{i,z^2,\uparrow} \  c^\dagger_{i,z^2,\downarrow} \ c^\dagger_{i,x^2-y^2,\uparrow} \  c^\dagger_{i,x^2-y^2,\downarrow}$ \\
$\ c^\dagger_{i,yz,\uparrow} \  c^\dagger_{i,yz,\downarrow} \ c^\dagger_{i,xz,\uparrow} \  c^\dagger_{i,xz,\downarrow} \ c^\dagger_{i,xy,\uparrow} \  c^\dagger_{i,xy,\downarrow}$).




\subsection{Hopping parameters from density functional theory}
\begin{figure}
\includegraphics[angle=0,width=\linewidth]{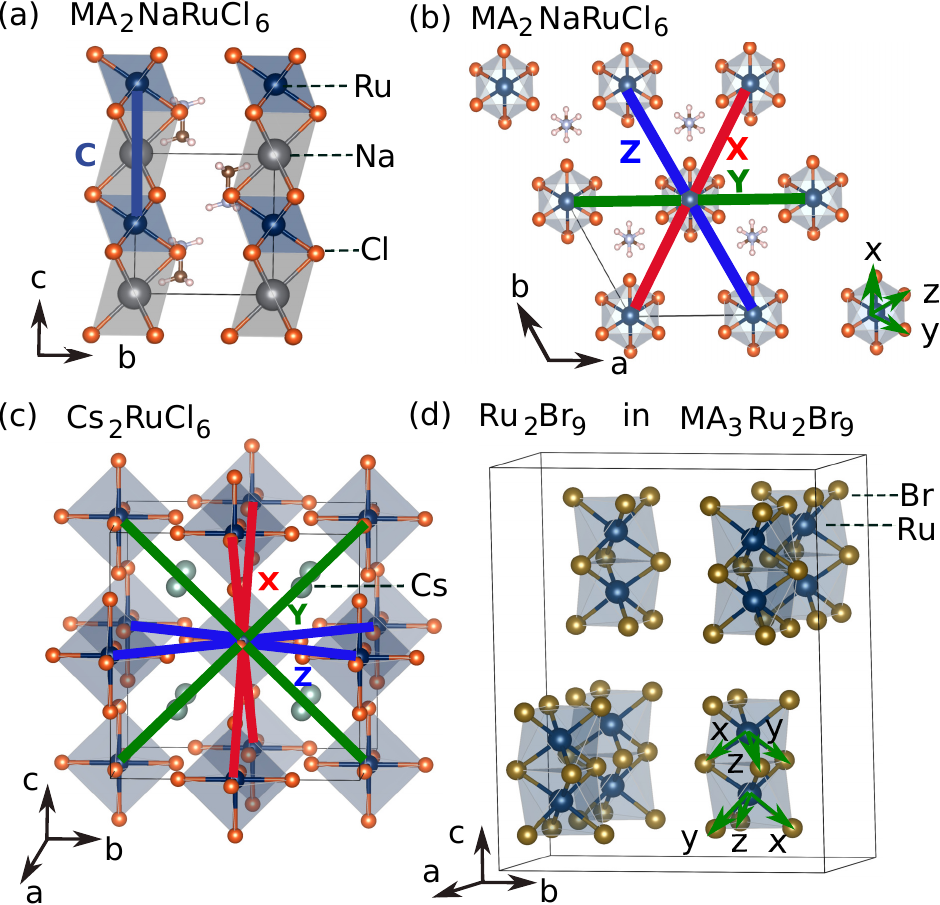}
\caption{Crystal structure of \NRC~ along (a) $bc$- and (b) $ab$-planes. (c) Crystal structure of \CsR.  The three different types of bonds $X$, $Y$, and $Z$, are shown in red, green and blue, respectively. (d) Ru$_2$Br$_9$ face-shared bioctahedral dimer in \RB. Green arrows denote local coordinates. 
}
\label{fig:structure}
\end{figure}
        \begin{table}[]
    \centering\def\arraystretch{1.1}
       \caption{Parameters for crystal-field splitting and hopping (meV) in Eq.\ref{eq:cf}-\ref{eq:dimer} extracted from DFT calculations for various materials.}
   \begin{ruledtabular}
 \begin{tabular}{lrrrrrr}    
\multicolumn{7}{c}{$d^5$ \NRC}\\
\hline
$\Delta $ & $t_C$& $ t_C^{\prime} $ & $t_1$ & $t_2$ & $t_3$ & $t_4$ \\
 -10.0         & -4.1 & -18.0 & 2.3 & -7.4 & -39.1 & 0.6 \\
\hline
\multicolumn{7}{c}{$d^5$ \RB}\\
\hline
$\Delta_1$ & $\Delta_2$ & $\Delta_3$ & $t_a$ & $t_b$ & $t_a^\prime$ & $t_b^\prime$\\
-99.0 & -48.9 & -59.6 & -336.4 & -338.6 & -309.6 & -250.1 \\
\end{tabular}
\end{ruledtabular}
~\\
\vspace{1mm}
\centering $d^4$ and $d^2$\\
\vspace{1mm}
\begin{ruledtabular}
 \begin{tabular}{lrrrr}   
&   Cs$_2$RuCl$_6$ & Cs$_2$WCl$_6$ & Rb$_2$WCl$_6$ & MA$_2$WCl$_6$\\
 \hline
$\Delta$       & 0 & 0     &  0    &  1.8     \\
$t_1$      & 7.3 & 11.5  &  3.9  &  1.9     \\
$t_2$     & -8.3 & 9.6   & -9.6  & -12.9    \\
$t_3$      & -46.6 & -38.4 & -61.4 & -62.2   \\
$t_4$     & 0 & 0     &  0    &  0.6         \\
\end{tabular}

\end{ruledtabular}
\label{tab:hopfull}
\end{table}

The hopping parameters for the multiorbital Hubbard models were extracted using the Wannier function projection formalism proposed in Ref.~\cite{aichhorn2009,Foyevtsova2013,ferber2014} from the electronic structure of density functional theory with the full-potential-linearized-augmented-plane-wave basis (LAPW) as implemented in WIEN2k~\cite{Wien2k}. We employed the Perdew-Burke-Ernzerhof generalized gradient approximation~\cite{Perdew1996} as exchange-correlation functional. For the calculations we considered  a k-mesh of 300 {\bf k} points in the first Brillouin zone and the parameter RK$_{\rm max}$ was set to 7 (3) for systems without (with) hydrogens, respectively. 
We consider first the case of a $d^5$ system \NRC~ in Fig.~\ref{fig:structure} (a)-(b). We label the nearest-neighbor Ru-Ru bonds along the $c$ axis as $C$ bonds, and as $X$, $Y$, $Z$ bonds the next-nearest neighbor Ru-Ru bonds which lie in the $ab$ plane forming a Ru triangular lattice. In the $t_{2g}$ basis, the tetragonal distortion of the octahedron is zero while there is a small trigonal distortion $\Delta$. In the $4d$-based hybrid vacancy-ordered double perovskites \CsR~[see Fig.~\ref{fig:structure} (c)], each Ru has four nearest neighbor bonds along $ac$, $bc$, and $ab$ planes, labeled as $X$, $Y$, and $Z$ bonds, respectively. The local coordinates are set to be the same as the global $a$, $b$, $c$ axes.  Different from \CsR~ and Rb$_2$WCl$_6$, in \MAW, the W form a triangular lattice in the $ab$ plane, similar to \NRC. 
The crystal field $\mathbf{\Delta}_i$ for \NRC, \MAW, \CsR, and \CsW~are written as:
\begin{align}
\mathbf{\Delta}_i = \left(\begin{array}{ccc} 0& \Delta &\Delta \\ \Delta &0&\Delta \\ \Delta & \Delta & 0 \end{array} \right),
\label{eq:cf}
\end{align}
The hopping integrals for the nearest neighbour C-bond [Fig.~\ref{fig:structure} (a)] in \NRC~are given by:
\begin{align}
\mathbf{T}^{C} = \left(\begin{array}{ccc} t_C & t^{\prime}_C & t^{\prime}_C \\ t^{\prime}_C & t_C & t^{\prime}_C \\ t^{\prime}_C & t^{\prime}_C & t_C \end{array} \right).
\end{align}
The hopping integrals for the nearest neighbour Z-bond in \NRC, \MAW, \CsR, and \CsW~are written as:
\begin{align}
\mathbf{T}^{Z} = \left(\begin{array}{ccc} t_1 & t_2 & t_4 \\ t_2 & t_1 & t_4 \\ t_4 & t_4 & t_3 \end{array} \right).
\end{align}
While X and Y bonds are given by:
\begin{align}
\mathbf{T}^{X} = \left(\begin{array}{ccc} t_3 & t_4 & t_4 \\ t_4 & t_1 & t_2 \\ t_4 & t_2 & t_1 \end{array} \right),
\end{align}
\begin{align}
\mathbf{T}^{Y} = \left(\begin{array}{ccc} t_1 & t_4 & t_2 \\ t_4 & t_3 & t_4 \\ t_2 & t_4 & t_1 \end{array} \right).
\end{align}

\begin{figure}
\center
\includegraphics[angle=0,width=\linewidth]{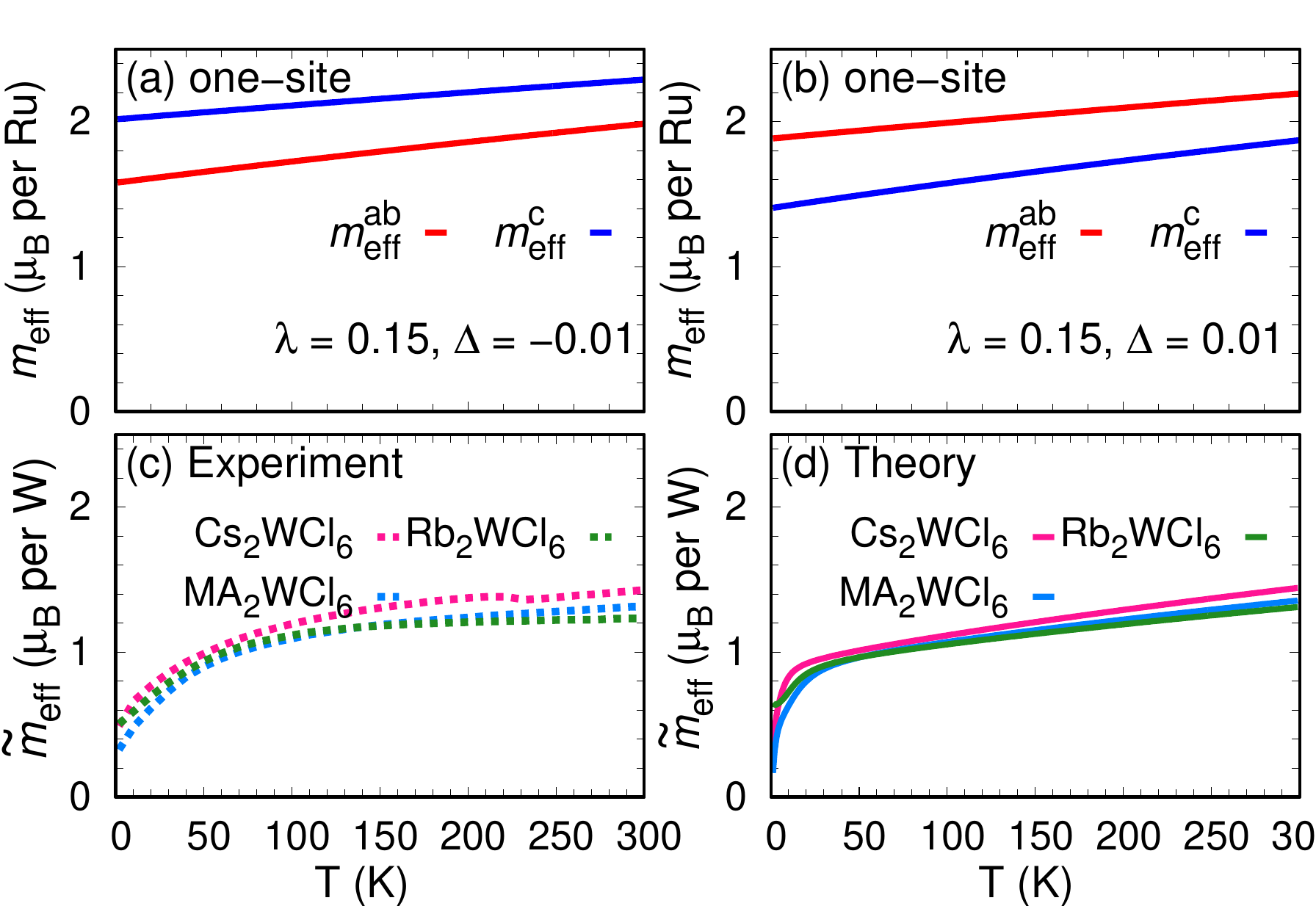}
\caption{Temperature dependence of the magnetic moments for in-plane direction ($m^{ab}_{\rm eff}$) and out-of-plane direction ($m^{c}_{\rm eff}$) in $d^5$ system \NRC~for (a) $\Delta$ = -0.01 eV and (b) $\Delta$ = 0.01 eV with $\lambda$ = 0.15 eV. (c), (d) Comparison of experimental~\cite{Vishnoi2021, Morgan2023} and calculated magnetic moments for \CsW, Rb$_2$WCl$_6$, and \MAW.}
\label{fig:momsup}
\end{figure}
For \RB~ shown in Fig.~\ref{fig:structure} (d), the Ru-Ru bond distances are around 2.4 - 2.5 \AA, forming  Ru-Ru bonding. Different from the case of inorganic compounds including M$_2$O$_9$ (M = Ir, Ru) dimers~\cite{Li2020}, where the local $D_{3h}$ point group symmetry within each dimer implies only one onsite trigonal term $\Delta$ and two types of hopping integrals: diagonal and off-diagonal in the local M-O  $x$, $y$, $z$ coordinates, the three Ru-Br bond-lengths in \RB~ are different, leading to splittings of the onsite crystal field
\begin{align}
\mathbf{\Delta}_i = \left(\begin{array}{ccc} 0& \Delta_2 &\Delta_1 \\ \Delta_2 &0&\Delta_1 \\ \Delta_1 & \Delta_1 & \Delta_3 \end{array} \right).
\end{align}
The hopping matrix intra the dimers for \RB~are written as:
\begin{align}
\mathbf{T} = \left(\begin{array}{ccc} t_a & t_b & t_b^{\prime} \\ t_b & t_a & t_b^{\prime} \\ t_b^{\prime} & t_b^{\prime} & t_a^{\prime} \end{array} \right).
\label{eq:dimer}
\end{align}
The corresponding hopping parameters are displayed in Table~\ref{tab:hopfull}.

\subsection{Direction-dependent magnetic moments for $d^5$ system and average magnetic moments for $d^2$ materials}

$\Delta$ strongly affects the direction-dependent magnetic moments $m^{ab}_{\rm eff}$ and $m^{c}_{\rm eff}$ as shown in Fig.~\ref{fig:momsup} (a) and (b) for \NRC. Our method also reproduces the experimental observations for other $d^2$ systems Rb$_2$WCl$_6$ and \MAW~[see Fig.~\ref{fig:momsup} (c), (d)]. 

\setlength{\bibsep}{0.0cm}

\end{document}